\documentclass[reviewcopy]{elsarticle}

\usepackage[reviewcopy]{adndt}
\usepackage{tabularx}
\usepackage{mathptmx}
\usepackage{times}
\usepackage{lscape}
\usepackage{amsmath}
\usepackage{amssymb}
\usepackage[dvipdfm,colorlinks=true,linkcolor=blue,urlcolor=blue,citecolor=blue]{hyperref}
\biboptions{square,sort&compress}
\bibpunct[]{[}{]}{,}{n}{}{;}
\newcommand{\ie}{\textit{i.e.,}}
\newcommand{\eg}{\textit{e.g.,}}
\newcommand{\code}[1]{\texttt{\footnotesize #1}}

\begin{document}

\begin{frontmatter}

\journal{Atomic Data and Nuclear Data Tables}

\title{Compilation of isomeric ratios of light particle induced nuclear reactions}

\author[One,Two]{A. Rodrigo}
\ead{E-mail: a.rodrigos@alumnos.upm.es}

\author[One]{N. Otuka\corref{cor1}}
\ead{E-mail: n.otsuka@iaea.org}
\cortext[cor1]{Corresponding author.}

\author[Three]{S. Tak\'{a}cs}

\author[One]{A.J. Koning}

\address[One]{Nuclear Data Section, Division of Physical and Chemical Sciences, Department of Nuclear Sciences and Applications, International Atomic Energy Agency, A-1400 Wien, Austria}

\address[Two]{Department of Energy Engineering (Division of Nuclear Engineering), Universidad Polit\'{e}cnica de Madrid, 28006 Madrid, Spain}

\address[Three]{Institute for Nuclear Research (ATOMKI), 4026 Debrecen, Hungary}

\date{12.01.2023} 

\begin{abstract}  
Experimental isomeric ratios of light (A$\le$4) particle-induced nuclear reactions were compiled for the product nuclides having metastable states with half-lives longer than 0.1~sec.
The experimental isomeric ratio data were taken from the EXFOR library and reviewed.
When an experiment reports isomer production cross sections instead of isomeric ratios, the cross sections taken from the EXFOR library were converted to the isomeric ratios by us.
During compilation, questionable data (\eg preliminary data compiled in EXFOR in parallel with their final data, sum of isomer production cross sections larger than the total production cross sections) were excluded.
As an application of the new compilation, goodness-of-fit was studied for the isomeric ratios predicted by the reaction model code TALYS-1.96.
\end{abstract}

\end{frontmatter}



\newpage

\tableofcontents
\listofDtables
\listofDfigures
\vskip5pc

\section{Introduction}

A nucleus on an excited level formed as a reaction product is typically deexcited to the ground state promptly by a series of gamma-ray emissions.
However, this deexcitation may be delayed due to presence of a long-lived excitation level.
Such an excitation level is known as the metastable state, whose spin is usually not close to the spin of the ground state and it prevents immediate deexcitation to a lower level.
It may further undergo deexcitation by gamma-ray emission to a lower level (isomeric transition) and/or by $\alpha$/$\beta$-ray emission or electron conversion to a neighboring nuclide.
Detection of such radiation allows us to measure the production cross section of the metastable state.
Similarly, we can define the production cross section of the ground state, which corresponds to deexcitation of the reaction product decayed into the ground state without going through any metastable state.
If there is only one metastable state, the total production cross section $\sigma_t$ is related with the ground state production cross section $\sigma_g$ and metastable state production cross section $\sigma_m$ by $\sigma_t=\sigma_g+\sigma_m$.

The ratio of production cross sections such as $\sigma_m/\sigma_g$ or $\sigma_m/\sigma_t$ is known as the isomeric ratio.
From the view of theoretical reaction modelling, the isomeric ratio is related with the spin ($J$) dependence of the level density of the intermediate and final product nuclei.
This distribution has been theoretically modelled by $(2J+1)\exp[-(J+1/2)^2 /(2\sigma^2)]$ with the square of the distribution width $\sigma^2$ known as the spin cut-off parameter \cite{Bethe1937,Bloch1954}.
Huizenga and Vandenbosch formulated the relationship between $\sigma$ and the isomeric ratio~\cite{Huizenga1960}, and various attempts have been made to parameterize $\sigma$ by using experimental isomeric ratios.
Namely, compilation of experimental isomeric ratios contributes to better model description of the isomer production cross sections through adjustment of the spin cut-off parameters to reproduce the compiled isomeric ratios. 

The knowledge of the isomeric ratio is also important for nuclear technology.
For example, the $^{241}$Am(n,$\gamma$)$^{242}$Am isomeric ratio is important from the view of nuclear waste management.
This is because $^{241}$Am may be produced in fission energy systems by successive neutron captures and $\beta^-$ decay starting from $^{238}$U, and the long-lived metastable state $^{242m}$Am can be further transmuted into the heavier americium isotopes and finally to $^{244}$Cm~\cite{Wisshak1982} due to a large thermal neutron capture cross section of $^{242m}$Am~\cite{Shinohara1997}(1290$\pm$300~b~\cite{Mughabghab2018}).
The isomeric ratios of low-energy neutron-induced reaction products have been evaluated and compiled in File 9 of the ENDF-6 format~\cite{Trkov2018}, and are utilized in reactor burn-up calculation.
The isomeric ratios for production of some metastable states such as $^{99m}$Tc and $^{186m}$Re in nuclear reactions are also important from the view of medical isotope production~\cite{Tarkanyi2019,Tarkanyi2022}.
Accessibility to the experimental isomeric ratios is, therefore, important for both research of nuclear reactions and its application.

When the ground and metastable states are unstable and their activities are measurable, the isomeric ratio is related with the counts of the ground and metastable state decays $N_g$ and $N_m$ by
\begin{equation}
\frac{\sigma_g}{\sigma_m}=\frac{f_m}{f_g}\left[
\frac{N_gI_{\gamma m}\epsilon_m}{N_mI_{\gamma g}\epsilon_g}-p\frac{\lambda_g}{\lambda_g-\lambda_m}
\right]
+p\frac{\lambda_m}{\lambda_g-\lambda_m}
\label{eqn:Vanska}
\end{equation}
~\cite{Vanska1981}, where $N$, $I_\gamma$, $\epsilon$, $p$ and $\lambda$ are the number of gamma-rays counted, gamma emission probability, gamma-ray detection efficiency, isomeric transition probability and decay constant, respectively.
The time factor $f$ is defined by $f=[1-\exp(-\lambda t_i)]\exp(-\lambda t_c)[1-\exp(-\lambda t_m)]/\lambda$ with the irradiation time $t_i$, cooling time $t_c$ and measurement time $t_m$.
This equation does not require determination of the incident particle flux, which may be a major source of the uncertainty and error in determination of the production cross section.
Similarly, prediction of the isomeric ratio by a reaction model is free from the absolute normalization (\eg total reaction cross section constrained by the optical potential).
These facts show an advantage to do comparison between measurements and model predictions for the isomeric ratio rather than for the isomer production cross sections.

The experimental isomeric ratios of nuclear reaction products have been compiled in the EXFOR library by the International Network of Nuclear Reaction Data Centres (NRDC)~\cite{Otuka2014}.
The compiled data are included in database systems and disseminated to the end users by the data centres~\cite{Zerkin2018,Soppera2014,Otuka2005,Boboshin2001}.
However, the isomeric ratios compiled in EXFOR have not been fully utilized because they are published in various expressions (\eg $\sigma_m/\sigma_t$, $\sigma_m/\sigma_g$), and the EXFOR library compiles these ratios as they are published without unification of the expression.
When an experimentalist reports $\sigma_g$ and $\sigma_m$ without their ratios, the ratios are not compiled in the EXFOR library, and this also makes the experimental information on the isomeric ratio less accessible.

In the past, assignment of the ground and metastable states has not been done in a consistent manner in EXFOR since assignments may depend on the decay scheme referred to by the experiment (\eg the 69~min state of $^{110}$In, which was known as the ground state in the past but now considered as a metastable state).
However, this inconsistency was analysed and improved by the data centres in 2010s~\cite{Semkova2017}. 
Considering these situations of EXFOR, we decided to compile experimental isomeric ratios which are derived from but are more accessible than those in the EXFOR library.
In the following sections, we discuss procedure of compilation and its application to benchmark of the TALYS-1.96 reaction model code~\cite{Koning2012}.

\section{Procedure}

\subsection{Criteria of data selection}

We defined the scope of our compilation by the following criteria:
\begin{itemize}
\item Experimental isomeric ratios or production cross sections compiled in EXFOR as of 23 August 2022.
\item Data not superseded. (\ie preliminary data are excluded if their final data are also in EXFOR.)
\item Data measured with a monoenergetic photon, neutron, proton, deuteron, triton, helion, or alpha particle beam.
\item Data for production of nuclides having ground state and only one metastable state with its half-life longer than 0.1~sec.
\end{itemize}
The EXFOR data were extracted not directly from the original EXFOR files but from the X4Pro database~\cite{Zerkin2022}.

In the EXFOR library, the quantity of each dataset is expressed by a REACTION code.
For example, the REACTION code \code{(79-AU-197(N,3N)79-AU-195-M,,SIG)} expresses the $^{197}$Au(n,3n)$^{195m}$Au cross section.
The two codes \code{79-AU-195-M} and \code{,SIG} express the product nuclide and quantity, respectively.
The combinations of the reaction product and quantity within our scope are summarized in Table~\ref{tab:reaction}.
Note that the code \code{ELEM/MASS} indicates that the atomic and mass numbers of the reaction product are independent variables of the EXFOR dataset.
See Chapter 6 of EXFOR Formats Manual~\cite{Otuka2022} for more details about the EXFOR REACTION formalism.
The EXFOR library also compiles the ground state production cross section including partial feeding via isomeric transition from a metastable production cross section (\eg \code{34-SE-73-G,M+,SIG}) and the production cross section including feeding by decay of another nuclide (\eg \code{13-AL-27,CUM,SIG}).
Such datasets are not for direct use of isomeric ratio construction and were excluded in the present compilation.

Isomeric ratios of fission products are also excluded for all spontaneous fission datasets, majority of the neutron-induced fission datasets and some other fission datasets.
This is because they are compiled in EXFOR as fission product yield ratios \code{FY/RAT} rather than the cross section ratios \code{SIG/RAT}, and their REACTION coding rule is slightly different (\eg the code indicating partial feeding \code{M+} is not combined with \code{FY/RAT}).
The readers are reminded that compilation of experimental isomeric fission yield ratios has been recently published by the US National Nuclear Data Center (NNDC)~\cite{Sears2021}.
Sometimes we found an experiment showing completely different trend from the other experiments.
When appearance of such an outlier was due to a typo in the EXFOR library, we fixed it.
Otherwise, we included such outliers in the present compilation without exclusion.

We also sometimes meet an experiment reporting $\sigma_t$ smaller than $\sigma_m$, and such an experiment was excluded from our compilation unless it was resolved by communication with the experimentalist.
This is often due to presence of a ground state production cross section without clear indication of the state in the nuclide symbol in documentation.
For example,
we experienced this problem for the cross sections tabulated with not $^{148g}$Pm but $^{148}$Pm published by Lebeda et al.~\cite{Lebeda2012,Lebeda2014}, for which the author kindly confirmed that they are not the total but the ground state production cross sections, and we were able to keep them in our compilation.
The EXFOR datasets corrected and excluded in the above-mentioned procedures are summarized in NRDC technical memos~\cite{CP-D1058,CP-D1060}.
Another possible reason of the unexpected relation between $\sigma_g$ and $\sigma_m$ is due to large uncertainties in the cross sections (e.g., $^{85}$Rb(p,x)$^{84}$Rb cross sections of Kastleiner et al.\cite{Kastleiner2004}, where we see $\sigma_m > \sigma_t$ at some incident energies though their error bars overlap.).

E.A. Skakun et al. measured (p,n) and (p,$\gamma$) isomer productions below 10~MeV with the Kharkiv proton linear accelerator and published several times (\eg \cite{Skakun1987,Batii1988,Skakun1992}).
Though they are compiled as independent results in EXFOR, we assumed they are from the same measurements and selected one of them for compilation as summarized in NRDC technical memo~\cite{CP-D1065}.
Unfortunately the isomeric ratios in its final publication~\cite{Skakun1992} are compiled in EXFOR by digitization from the figure images and we did not adopt them.

A reference value (\eg monitor cross section, gamma emission probability) adopted by the experimentalist may be different from the currently recommended value.
We did not update the originally published data compiled in EXFOR during the present compilation except for the proton-induced activation cross sections published by Levkovskii~\cite{Levkovskii1991}, for which we renormalized the originally published cross sections and compiled in EXFOR A0511 by 192.8/252$\sim$0.77 where 252~mb is the $^{\mathrm nat}$Mo(p,x)$^{96}$Tc cross section at 30~MeV adopted by Levkovskii while 192.8~mb is the value recommended by an IAEA Coordinated Research Project~\cite{Hermanne2018}.

\subsection{Ground and metastable state assignments}

The ground and metastable state assignments may depend on the decay data adopted by the experimentalist.
During the comprehensive review and improvement of isomeric flagging in EXFOR performed in 2010s~\cite{CP-D0888,Semkova2017}, we followed the assignment seen in Nuclear Wallet Cards~\cite{Tuli2011}.

Some experimentalists do not consider a short-lived metastable state as an isomer.
For example, the first metastable state of $^{196m1}$Au (8.1~sec) is usually not detectable in an activation measurement designed for detection of $^{196g}$Au (6.2~d) and $^{196m2}$Au (9.6~hr) activities.
Consequently, an experimentalist may report their $^{196m2}$Au production cross sections just as $^{196m}$Au production cross sections, which may be wrongly entered in EXFOR as \code{79-AU-196-M,,SIG} though this must be \code{79-AU-196-M2,,SIG}.
In order to exclude such a dataset compiled with improper isomeric flagging, the reaction product code of each EXFOR dataset was checked against NUBASE~\cite{Kondev2021}, and the dataset was excluded when NUBASE defines two or more metastable states or no metastable state for the product nuclide.
Typical examples of such nuclides are (1) $^{124m2}$Sb 20~min states (denoted as $^{124m}$Sb in the literature(\eg \cite{Paul1953,Qaim1968}) and (2) $^{30m}$Al 72.5~sec state, whose production cross sections were reported in the past~(\eg \cite{Grench1971,Schantl1970}) but this state is currently unknown.

\subsection{Conversion of cross sections to isomeric ratios}

After extraction of the EXFOR datasets within our scope and filtered by the above-mentioned procedures, we converted the extracted data to the isomeric ratios $\sigma_m/\sigma_t$.
When an experimental work does not provide an isomeric ratio in EXFOR but provide at least two of $\sigma_g$, $\sigma_m$ and $\sigma_t$ at the same incident energy, we converted them to $\sigma_m/\sigma_t$ for compilation.
When all these three types of the cross sections are available, we did not use $\sigma_g$.
If an experiment does not provide any pair of the cross sections at the same incident energy, we simply discarded the experiment.

An experiment may report two or more data points at the same incident energy.
%
%
When an average value from several measurements is reported, we adopted it while discarded the individual results.
For example, Meierhofer et al.~\cite{Meierhofer2010} reports 6 $\sigma_t$ and 12 $\sigma_m$ values for the $^{74}$Ge(n,$\gamma$)$^{75}$Ge reaction at the thermal energy, and one may construct 72 $\sigma_m/\sigma_t$ values from various combinations of $\sigma_t$ and $\sigma_m$.
However, they also report the average of the $\sigma_t$ and $\sigma_m$ values and we adopted only these averages to obtain a single $\sigma_m/\sigma_t$ value from this measurement.
When the authors report only individual results without their averages, we compiled the isomeric ratios derived from all combinations of the cross sections, and tabulated them with a flag for caution.

Filatenkov et al. performed systematic measurements of neutron activation cross sections and documented their results around 2000~\cite{Filatenkov1999,Filatenkov2001}.
Later the cross sections were revised with the updated reference data (\eg decay data) and published in 2016~\cite{Filatenkov2016}.
During our compilation, we found some isomeric ratios in the two original reports are not seen in the 2016 report even in revised forms though the corresponding cross sections are there.
We adopted the isomeric ratios derived from the cross sections published in the 2016 report rather than the isomeric ratios published in the original reports.~(c.f. \cite{CP-D1061,CP-D1062}).
Similarly, we found that an isomeric ratio derived from the high energy (above 660~MeV) cross sections measured by A.R.Balabekyan et al. at JINR (\eg \cite{Alexandryan1996,Aleksandryan2002,Balabekyan2006,Danagulyan2013,Balabekyan2014,Balabekyan2016}) and compiled in an EXFOR entry is often very close to an isomeric ratio compiled in another EXFOR entry. We carefully identified such pairs to avoid appearance of the isomeric ratios from the same experiment twice~(c.f. \cite{CP-D1066}).

\subsection{Uncertainty}

The EXFOR library may provide several types of the uncertainties such as the total uncertainty (\code{ERR-T}), statistical uncertainty (\code{ERR-S}), total systematic uncertainty (\code{ERR-SYS}), partial uncertainty (\code{ERR-1}, \code{ERR-2} etc.) or uncertainty without further specification (\code{DATA-ERR}).
When several of them are in EXFOR, we always selected the largest one in our tabulation.
When an isomeric ratio was derived from the respective cross sections, we propagated the uncertainties in the cross sections to the isomeric ratio assuming that the uncertainties in the cross sections are independent.
This may overestimate the actual uncertainty since a partial uncertainty (\eg uncertainty in the incident particle flux) may be shared in both cross sections and cancelled when they are converted to the isomeric ratio.

The isomeric ratio plus (minus) its uncertainty in our compilation is sometimes higher (lower) than 1 (0).
There are a few such ratios directly taken from the original publication (\eg, $^{75}$As(n,p)$^{75}$Ge isomeric ratio in Ref.~\cite{Bormann1966}) but the majority of them are $\sigma_m/\sigma_t$ values derived by us.
Such values are flagged in the main table for caution.

\section{Results}

Table~\ref{tab:statistics} summarizes the number of reactions and isomeric ratios for each projectile.
Very few photon-induced reaction isomeric ratios were found for inclusion in the current compilation.
This is because usually photoactivation isomeric ratio measurements are done with bremsstrahlung photon sources, which are not monoenergetic and not for our compilation.

It is not an intention of this article to discuss various findings in individual cases.
Nevertheless, we discuss activation measurements of two reactions just to demonstrate what kind of discussion we can do based on the new compilation.

\subsection{$^{93}$Nb(n,$\alpha$)$^{90}$Y}
\begin{figure}[ht!]
\centering
\includegraphics[width=.7\linewidth]{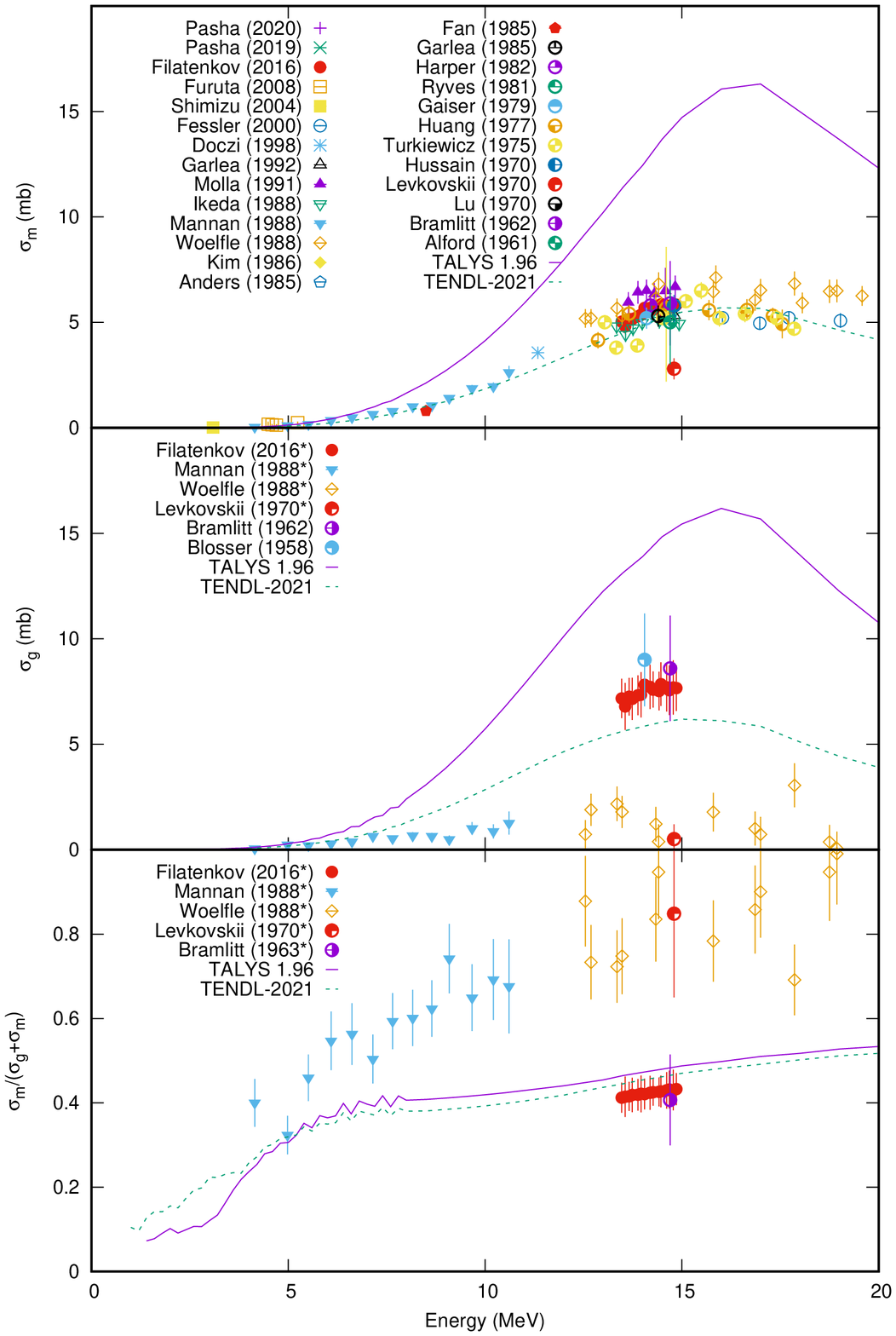}
\caption{
$^{93}$Nb(n,$\alpha$)$^{90}$Y metastable (top) and ground (middle) state production cross sections and isomeric ratios (bottom).
The asterisk after the year indicates that the values are not reported by the experimentalists but derived from the original values compiled in EXFOR (\eg isomeric ratio derived from isomer production cross sections).
In addition to the experimental data~\cite{Filatenkov2016, Pasha2020, Pasha2019, Furuta2008, Shimizu2004, Fessler2000, Doczi1998, Garlea1992, Molla1991, Ikeda1988, Mannan1988, Woelfle1988, Kim1986, Pepelnik1986, Fan1985, Garlea1985, Harper1982, Ryves1981, Gaiser1979, Huang1977, Turkiewicz1975, Hussain1970, Levkovskii1970, Lu1970, Bramlitt1963, Alford1961,Blosser1958} extracted from the present compilation,
the corresponding data predicted by TALYS-1.96 and evaluated for the TENDL-2021 library~\cite{Koning2019} are plotted in curves.
}
\label{fig:93Nb}
\end{figure}

Figure~\ref{fig:93Nb} shows the $^{93}$Nb(n,$\alpha$)$^{90}$Y isomeric ratios as well as the ground and metastable state production cross sections~\cite{Filatenkov2016, Pasha2020, Pasha2019, Furuta2008, Shimizu2004, Fessler2000, Doczi1998, Garlea1992, Molla1991, Ikeda1988, Mannan1988, Woelfle1988, Kim1986, Pepelnik1986, Fan1985, Garlea1985, Harper1982, Ryves1981, Gaiser1979, Huang1977, Turkiewicz1975, Hussain1970, Levkovskii1970, Lu1970, Bramlitt1963, Alford1961,Blosser1958}.
The 3.2~hr metastable state has two intense gamma lines at 203.53 and 479.51~keV~\cite{Basu2020} and measurement of its production cross section is straightforward.
On the other hand, the 64~hr ground state does not have such a suitable gamma line, and it makes measurement of the ground state production cross section difficult.
Filatenkov carefully determined the isomeric ratio by decay-curve analysis for an energetic ($E_{\mathrm{max}}$=2280~keV) $\beta^-$-ray by a HPGe detector considering the fact that there are few other reaction products and they do not emit $\gamma$-rays at the high energy region.
See Sect. 2.7.2 of Ref.~\cite{Filatenkov2016} for more details.
The isomeric ratios reported by Filatenkov are lower than the majority of the isomeric ratios published by others but consistent with the prediction by TALYS.
On the other hand, the TALYS calculation with the same default parameters underestimates both ground and metastable state production cross sections.

\subsection{$^{197}$Au(d,2n)$^{197}$Hg}
\begin{figure}[ht!]
\centering
\includegraphics[width=.7\linewidth]{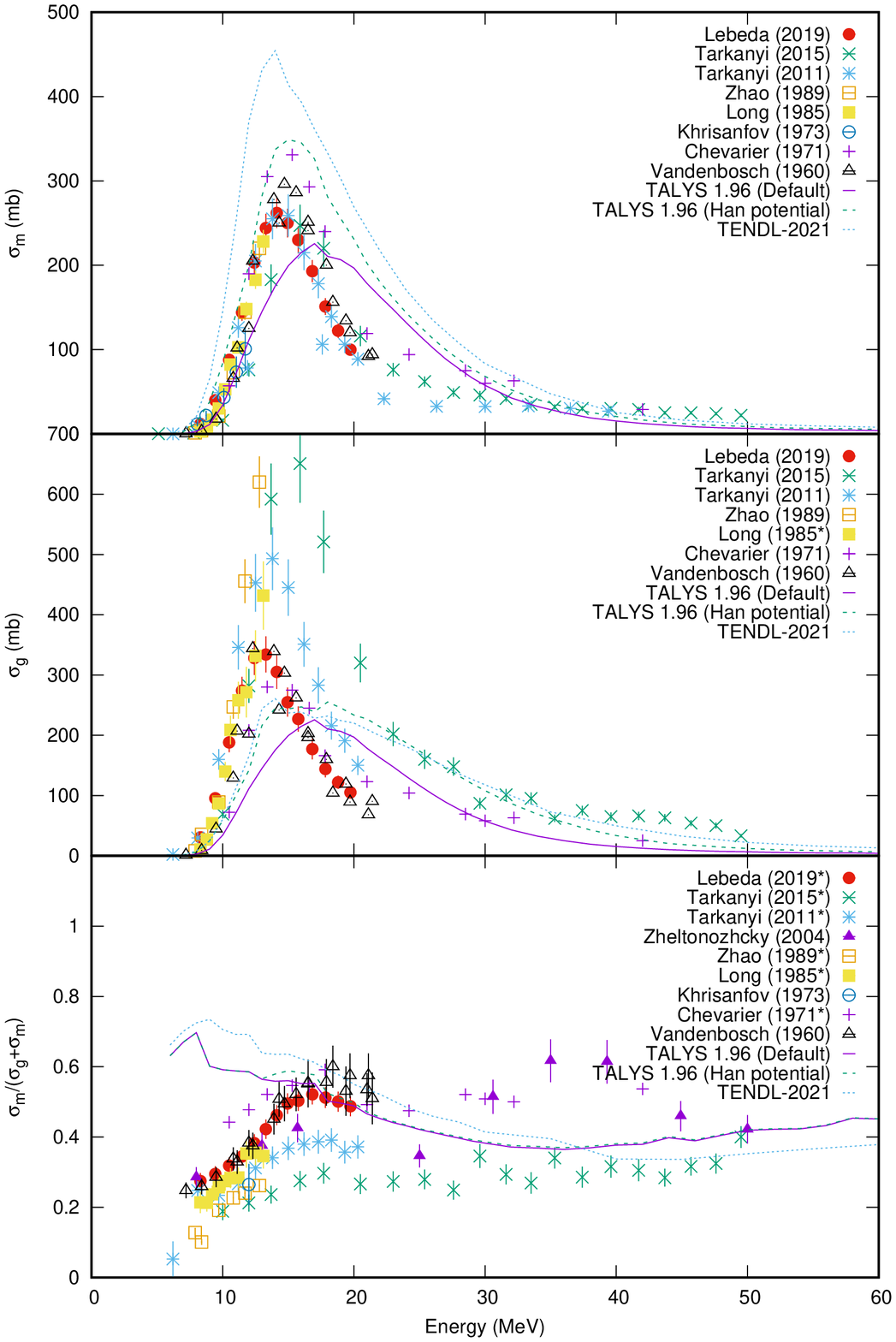}
\caption{
$^{197}$Au(d,2n)$^{197}$Hg metastable (top) and ground (middle) state production cross sections and isomeric ratios (bottom).
The asterisk after the year indicates that the values are not reported by the experimentalists but derived from the original values compiled in EXFOR (\eg isomeric ratio derived from isomer production cross sections).
In addition to the experimental data~\cite{Lebeda2019,Tarkanyi2015,Tarkanyi2011,Zheltonozhcky2004,Zhao1989,Long1985,Khrisanfov1973,Chevarier1971,Vandenbosch1960} extracted from the present compilation,
the corresponding data predicted by TALYS-1.96 (with the default deuteron potential and the potential proposed by Han et al.~\cite{Han2006}) and evaluated for the TENDL-2021 library~\cite{Koning2019} are plotted in curves.
}
\label{fig:197Au}
\end{figure}

Figure~\ref{fig:197Au} shows the $^{197}$Au(d,2n)$^{197}$Hg isomeric ratios as well as the ground and metastable state production cross sections~\cite{Lebeda2019,Tarkanyi2015,Tarkanyi2011,Zheltonozhcky2004,Zhao1989,Long1985,Khrisanfov1973,Chevarier1971,Vandenbosch1960}.
Not only the 24~hr metastable state but also the 64~hr ground state have characteristic gamma-rays, and it is possible to measure the production cross sections of both states in principle.
Similar to the $^{93}$Nb(n,$\alpha$)$^{90}$Y case, however, the ground state production cross sections are more scattered than the metastable state production cross sections in the literature.
Furthermore, we see several groups in the energy dependence of the isomeric ratios.

When the half-life of the ground state is longer than the half-life of the metastable state and the measurement was done after long cooling time allowing complete decay of the co-produced metastable state to the ground state (\ie $t_c \gg 1/\lambda_m$), the cross section derived from the measurement of the ground state activity $\sigma_c = N_g / (f_g I_{\gamma g} \epsilon_g n \phi)$ with the sample areal density $n$ and beam flux $\phi$ is sometimes assumed to be $\sigma_c \sim \sigma_g + p \sigma_m$.
This leads to 
\begin{equation}
\sigma_g \sim \sigma_c-p\sigma_m
\label{eqn:wrong}
\end{equation}
for determination of $\sigma_g$ from $\sigma_c$ and $\sigma_m$.
Below we demonstrate that this equation is valid only when $\lambda_m \gg \lambda_g$ or $\sigma_m\ll\sigma_g$.

Since $\sigma_m = N_m/(f_m I_{\gamma m} \epsilon_m n \phi)$, Eq.~(\ref{eqn:Vanska}) can be rewritten to 
\begin{equation}
\sigma_c = \sigma_g + p \left(\frac{f_m}{f_g}\frac{\lambda_g}{\lambda_g-\lambda_m}-\frac{\lambda_m}{\lambda_g-\lambda_m} \right) \sigma_m.
\end{equation}
If $t_c \gg 1/\lambda_m$, $\exp(-\lambda_m t_c)/\exp(-\lambda_g t_c) \to 0$, namely $f_m/f_g \to 0$ and
\begin{equation}
\sigma_c \to \sigma_g + p \frac{\lambda_m}{\lambda_m-\lambda_g}\sigma_m.
\end{equation}
Therefore, one can determine $\sigma_g$ after long cooling by
\begin{equation}
\sigma_g \sim \sigma_c -p\frac{\lambda_m}{\lambda_m-\lambda_g}\sigma_m.
\label{eqn:good}
\end{equation}
in general as long as $\lambda_m > \lambda_g$.

It follows from Eq.~(\ref{eqn:good}) that Eq.~(\ref{eqn:wrong}) is valid only when (1) $\lambda_m \gg \lambda_g$ or (2) $\sigma_m \ll \sigma_g$, and use of Eq.~(\ref{eqn:wrong}) adds an extra term $p[\lambda_m/(\lambda_m-\lambda_g)-1]\sigma_m$ to the actual ground state production cross section in general.
As $p[\lambda_m/(\lambda_m-\lambda_g)-1] \sim 0.53$ and $\sigma_m$ is not negligible for the $^{197}$Au(d,2n)$^{197}$Hg reaction, some experiments showing high $\sigma_g$ values in Fig.~\ref{fig:197Au} may include this extra term.
We notice that similar problems may occur in pairs of the metastable and ground states having close half-lives (\eg $^{198}$Tl, $^{198}$Au) and we wish our compilation will contribute to discussion on this problem.

\section{Application}

Global test of reaction model codes is an immediate application of the newly prepared isomeric ratio table.
We can easily check reaction model codes from the goodness-of-fit of outputs obtained using the new isomeric ratio table.
As an example, we calculated isomeric ratios by TALYS-1.96 with the default parameter sets but varying the spin cut-off parameter.
The spin cut-off parameter used in TALYS is
\begin{equation}
\sigma^2=R\frac{\tilde{a}}{a}\frac{I_\mathrm{rig}T}{\hslash^2}
\label{eqn:scm1}
\end{equation}
in default setting (``spincutmodel 1"), where $I_\mathrm{rig}$ is the rigid body moment of inertia, $T$ is the nuclear temperature, and $a$ and $\tilde{a}$ are the level density parameter and its high excitation energy limit, respectively.
In ``spincutmodel 2", this is simplified to 
\begin{equation}
\sigma^2=R\frac{I_\mathrm{rig} T}{\hslash^2}\equiv\frac{I_\mathrm{eff} T}{\hslash^2}
\label{eqn:scm2}
\end{equation}
by omitting the shell effect factor $a/\tilde{a}$.
$R$ is an adjustable parameter in TALYS.\footnote{$R$ may be specified by \code{Rspincut} (nuclide independent) or \code{s2adjust} (nuclide dependent) in TALYS-1.96.}
The parameter $\eta=I_\mathrm{eff}/I_\mathrm{rig}$ seen in the literature~\cite{Sudar2018,Avrigeanu2002,Dilg1973} is equal to $R$ in Eq.~(\ref{eqn:scm2}).
To see an appropriate choice of R, we calculated the isomeric ratios of all reactions in the present compilation from 1~eV (neutron-induced reactions) or 1~MeV (other reactions) to 200~MeV with the energy grids hardwired in TALYS.
The $R$ values were varied between 0.1 and 1.5, and calculations were done with both spin cut-off parameter models.

\begin{figure}[ht!]
\centering
\includegraphics[width=.7\linewidth,angle=-90]{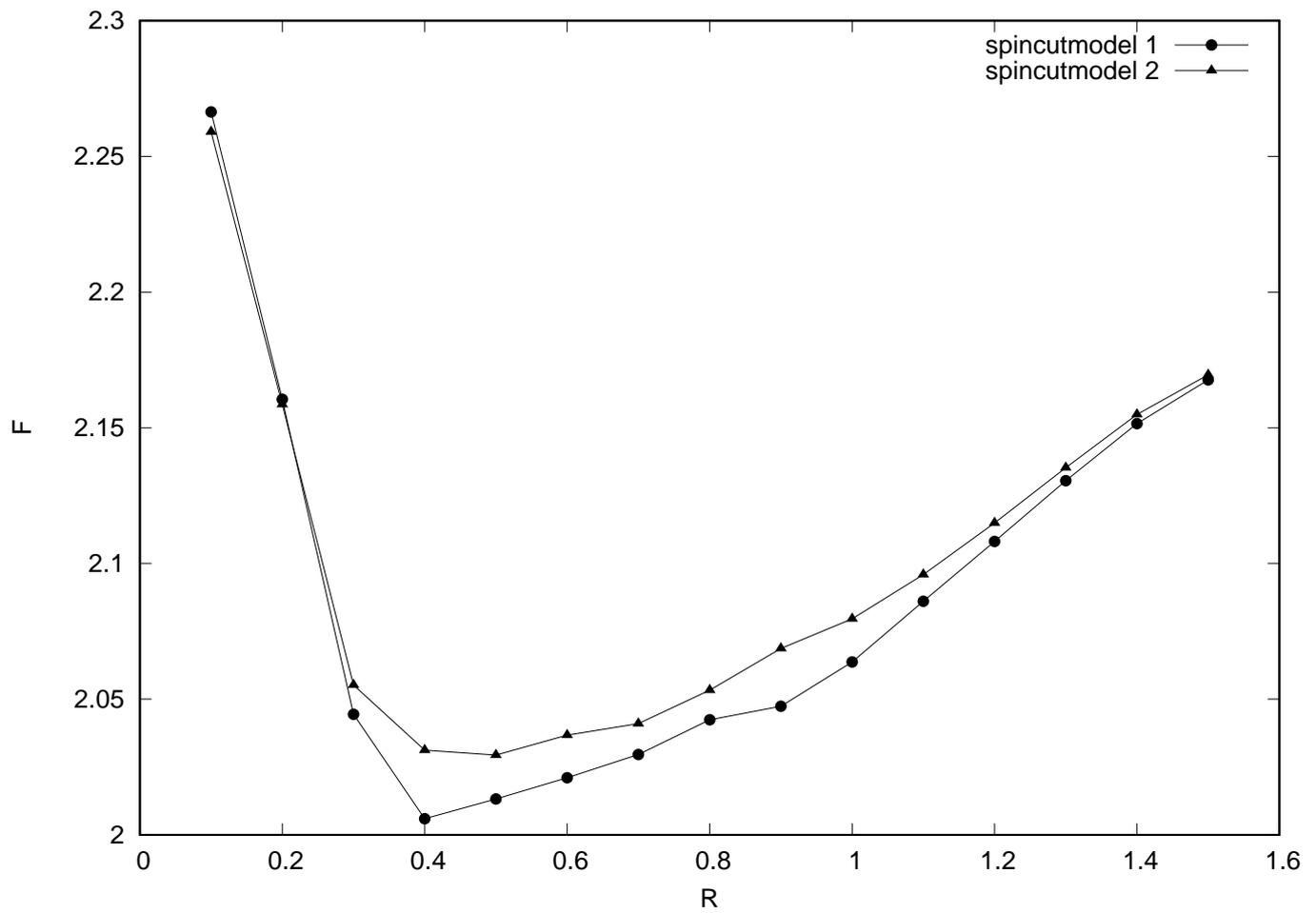}
\caption{
$R$ dependence of the $F$-value for the isomeric ratios predicted by TALYS-1.96 with two spin cut-off models.
}
\label{fig:fom}
\end{figure}

For $n \sim 12,000$ experimental isomeric ratios compiled in the present work with their uncertainties and the isomeric ratios predicted by TALYS, we calculated the $F$-value~\cite{Goriely2018}:
\begin{equation}
F=\exp\sqrt{\frac{1}{n}\sum_{i=1}^n\left[\ln\left(\frac{r_{i,\mathrm{cal}}}{r'_{i,\mathrm{exp}}}\right)\right]^2}
\end{equation}
with
\begin{equation}
r'_{i,\mathrm{exp}}=\left\{
\begin{array}{ll}
r_{i,\mathrm{exp}}- \Delta r_{i,\mathrm{exp}} & \mathrm{if~} r_{i,\mathrm{cal}}< r_{i,\mathrm{exp}} -\Delta r_{i,\mathrm{exp}} \\
r_{i,\mathrm{exp}}+ \Delta r_{i,\mathrm{exp}} & \mathrm{if~} r_{i,\mathrm{cal}}> r_{i,\mathrm{exp}} +\Delta r_{i,\mathrm{exp}} \\
1                                               &\mathrm{otherwise}
\end{array}
\right. ,
\end{equation}
where $r_{i,\mathrm{exp}}$ and $\Delta r_{i,\mathrm{exp}}$ are the $i$th isomeric ratio in our compilation and its uncertainty, and $r_{i,\mathrm{cal}}$ is the corresponding isomeric ratio predicted by TALYS.
Figure~\ref{fig:fom} shows $R$ dependence of the $F$-value.
This figure suggests that the best fit is obtained when the spin cut-off parameter is reduced to $\sim$40\% ($\sim$50\%) of its default value $R=1$ when using the ``spincutmodel 1" (``spincutmodel 2") setting.
For more sophisticated evaluation of isomeric ratios, $R$ must be adjusted for each nuclide individually.
For example, Sud\'{a}r et al.~\cite{Sudar2018} reports that the $\eta$ value shows strong mass dependence when it was adjusted for each nuclide separately.

\begin{figure}[ht!]
\centering
\includegraphics[width=.7\linewidth,angle=-90]{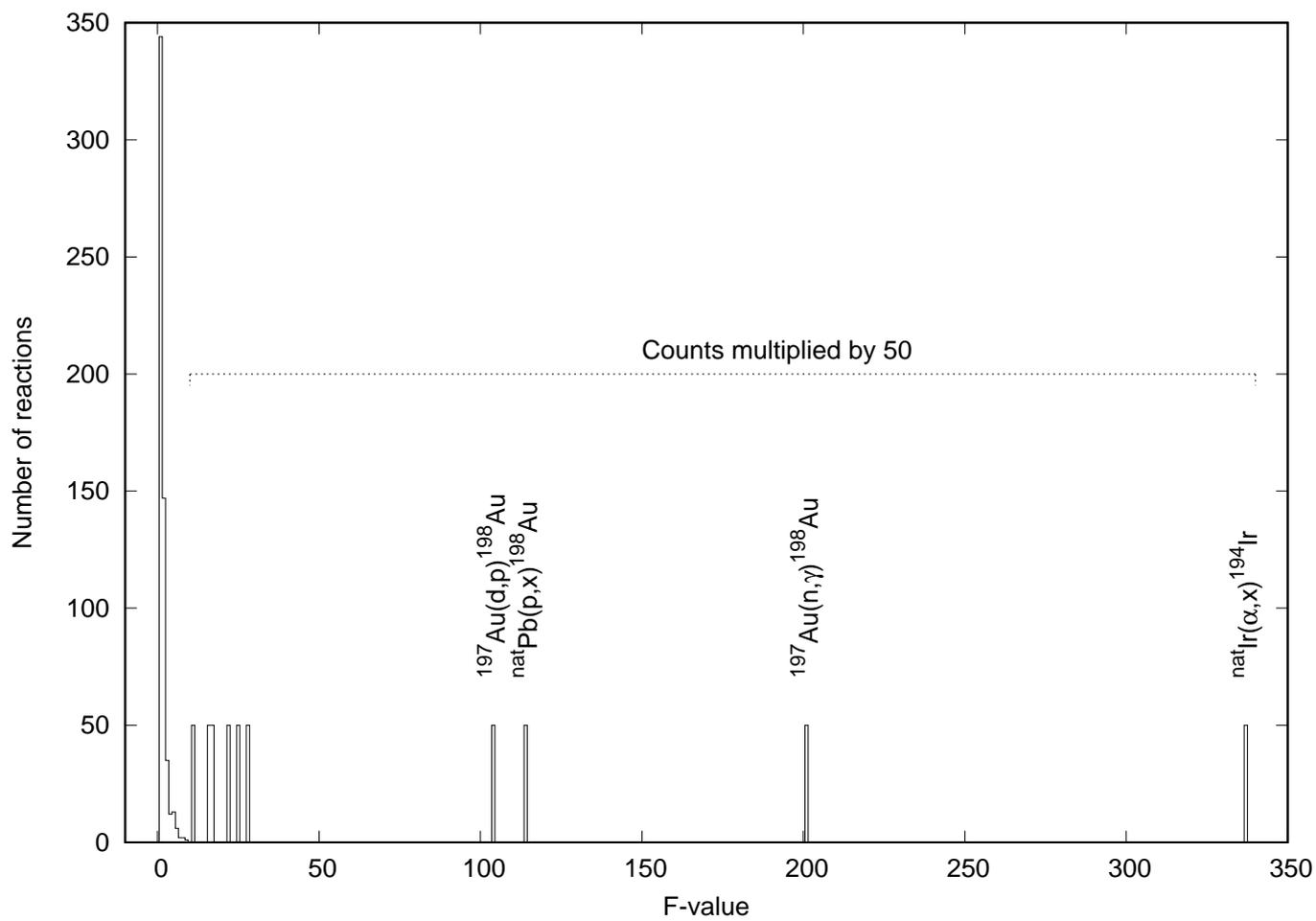}
\caption{
Distribution of $F$-value for the isomeric ratios predicted by TALYS-1.96 with default setting (``spincutmodel 1" and $R=1$).
}
\label{fig:fomdist}
\end{figure}

Figure~\ref{fig:fomdist} shows distribution of the $F$-value for prediction by TALYS-1.96 with default setting (``spincutmodel 1" and $R=1$).
Among four reactions getting high $F$-values, $^{197}$Au(n,$\gamma$)$^{198}$Au and $^\mathrm{nat}$Pb(p,x)$^{198}$Au could be difficult ones to get $F\sim 1$ since $\sigma_m/\sigma_t$ is very low ($\sim 10^{-3}$ or lower) for the former reaction, and the $F$-value is based on only one experimental $\sigma_m/\sigma_t$ value at very high energy (150~MeV) for the latter reaction.
On the other hand, we observe systematic deviations of experimental $\sigma_m/\sigma_t$ values from those predicted by TALYS-1.96 for $^\mathrm{nat}$Ir($\alpha$,x)$^{194}$Ir and $^{197}$Au(d,p)$^{198}$Au, for which the model predictions could be improved.
\section{Summary}

We extracted the experimental isomer production cross sections and isomeric ratios from the EXFOR library and compiled the isomeric ratios in the form of $\sigma_m/\sigma_t$.
Various mistakes in the EXFOR library and original publications were fixed during compilation.
Preliminary experimental results and experiments reporting unphysical $\sigma_m/\sigma_t$ values were discarded during compilation.
As an application of the newly created isomeric ratio table, we studied the spin cut-off parameter dependence of the goodness of fit for the isomeric ratios predicted by TALYS-1.96.

%

\section*{Data availability}
The table of the compiled isomer production cross sections and isomeric ratios in a plain text file is included in the supplemental material.
It is also available upon request to the authors by email or post.
Their graphical comparison with evaluated data libraries is under preparation for an IAEA report~\cite{Rodrigo2023}.


\ack
The authors are grateful to Viktor Zerkin for his special arrangement of a X4Pro SQLite database file dedicated to the present work.
We also wish to thank the anonymous reviewer for careful reading of the earlier version of the manuscript.
S\'{a}ndor Sud\'{a}r helped us to understand his adjustment of $\eta$ values with TALYS.
Ferenc Ditr\'{o}i, Gy\"{o}rgy Gy\"{u}rky, Alex Hermanne, Mayeen Uddin Khandaker, Ond\v{r}ej Lebeda, Rolf Michel, Haladhara Naik, Syed M. Qaim, Yury Titarenko and Sung Chul Yang helped us to understand the experimental data published by them.
We would like to thank Oscar Cabellos for encouraging us.
Last but not least, we appreciate the managers, compilers and programmers of the Nuclear Reaction Data Centres (NRDC) for maintenance and development of the EXFOR library.

\section*{Supplemental material}
The compiled cross sections and isomeric ratios in a plain text file and plots of the experimental cross sections and isomeric ratios along with those predicted by TALYS-1.96 with default setting can be found online at https://doi.org/10.1016/j.adt.2023.xxxxxx.
%


\clearpage

\section*{Figures}

\clearpage

\begin{table}
\caption{Combinations of reaction product and quantity considered in our compilation in EXFOR REACTION formalism. \code{Z-S-A} stands for a nuclide symbol (\eg \code{79-AU-198}). Note that the three underlines code strings are obsolete but still seen in old EXFOR entries.}
\label{tab:reaction}
\begin{center}
\begin{tabular}{ll}
\hline
REACTION (SF4-SF6)                   &Symbols            \\
\hline                              
\code{Z-S-A,,SIG}                    &$\sigma_t$         \\
\underline{\code{Z-S-A,IND,SIG}}&                        \\
\hline
\code{Z-S-A-G,,SIG}                  &$\sigma_g$         \\
\underline{\code{Z-S-A-G,M-,SIG}}    &                   \\
\hline
\code{Z-S-A-M,,SIG}                  &$\sigma_m$         \\
\hline                              
\code{Z-S-A-M/T,,SIG/RAT}            &$\sigma_m/\sigma_t$\\
\hline                              
\code{Z-S-A-G/T,,SIG/RAT}            &$\sigma_g/\sigma_t$\\
\hline                              
\code{Z-S-A-M/G,,SIG/RAT}            &$\sigma_m/\sigma_g$\\
\hline                              
\code{ELEM/MASS,,SIG}                &$\sigma_g$ (coded with \code{ISOMER=0}),    \\
\underline{\code{ELEM/MASS,IND,SIG}} &$\sigma_m$ (coded with \code{ISOMER=1}) or  \\
                                     &$\sigma_t$ (coded without \code{ISOMER}).   \\
\hline
\end{tabular}
\end{center}
\end{table}

\begin{table}
\caption{Number of reactions and data points of the isomeric ratios compiled in the present work.}
\label{tab:statistics}
\begin{center}
\begin{tabular}{lccccccc}
\hline
projectile &$\gamma$&    n&    p&    d&$^{3}$He & $\alpha$&total  \\
\hline
reactions  &       2&  186&  470&  127&       34&      143&   962 \\
\hline            
points     &       9& 2229& 4883& 1915&      624&     2653& 12313 \\
\hline
\end{tabular}
\end{center}
\end{table}
\datatables 
\newpage
\end{document}